\documentstyle[11pt] {article}

\title{Hierarchical Construction of Finite Diabatic Sets,
 By Mathieu Functions}
\author{
R. Englman$^{a,b}$, A. Yahalom$^b$ and M. Baer$^a$\\
$^a$ Department of Physics an Applied Mathematics,\\
Soreq NRC,Yavne 81810,Israel\\
$^b$ College of Judea and Samaria, Ariel 44284, Israel\\
e-mail: englman@vms.huji.ac.il; asya@ycariel.yosh.ac.il;\\
mmbaer@netvision.net.il;}

\begin{document}
\maketitle

\newcommand{\beq} {\begin{equation}}
\newcommand{\enq} {\end{equation}}
\newcommand{\ber} {\begin {eqnarray}}
\newcommand{\enr} {\end {eqnarray}}
\newcommand{\eq} {equation}
\newcommand{\eqs} {equations }
\newcommand{\mn}  {{\mu \nu}}
\newcommand{\sn}  {{\sigma \nu}}
\newcommand{\rhm}  {{\rho \mu}}
\newcommand{\sr}  {{\sigma \rho}}
\newcommand{\bh}  {{\bar h}}
\newcommand {\er}[1] {equation (\ref{#1}) }

\begin {abstract}
 An extension is given for the standard two component model of adiabatic,
 Born-Oppenheimer (BO)
electronic states in a polyatonic molecule, by use of Mathieu functions of
arbitrary order. The
curl or compatibility conditions for the construction of a diabatic set of
 states based on a finite-
dimensional subset of BO states are not satisfied exactly. It is
 shown, however, that, by successively adding higher order Mathieu functions
 to the BO set, the compatibility conditions are satisfied with increasingly
better accuracy.
We then generalize to situations in which the nonadiabatic couplings
(the dynamic corrections to the BO approximation)
 are small (though not necessarily zero) between
 a  finite-dimensional BO subset and the rest of the BO states.
We prove that approximate diabatic sets exist, with an error that is of
 the order of the square of the neglected nonadiabatic couplings.

\bigskip

PACS number(s):31.15.-p, 31.30.-i, 31.30.Gs, 31.70.-f

\bigskip
keywords: Born-Oppenheimer states, nonadiabatic coupling, Mathieu functions,
diabatic set, Yang-Mills field
\end {abstract}

\section { Background and Introduction, preceded by a Homage}
\label{BaI}

Some of the authors in the article have had the good fortune that their pathways
crossed with those of Per-Olov Lowdin, be this in Slater's Group at MIT, in
Menton or at the Sanibel Workshops. They keep the memory of an endowed and innovative
 researcher, as well as of a Scientist Statesman, who followed in the footsteps
 of another great Scandinavian, Niels Bohr.

\bigskip

The question of whether it is possible to effect in general an adiabatic
to diabatic transformation (ADT) has been recently reopened
\cite {BA},\cite {MB0},\cite{BLAAB}. One
formulation of the issues \cite{Sidis}-\cite{TK}
 is whether a strictly diabatic basis in which linear
combinations of a small number of Born-Oppenheimer electronic states are
chosen in such a way that the nonadiabatic coupling between the transformed
states vanishes exists.

In view of possible ambiguities let us define the terms:
\beq
\zeta_k \equiv \zeta_k ({\bf x_e},X_r)
\label{zet}
\enq
is a  solution of the Born-Oppenheimer wave-equation for the electronic part
(electronic coordinates ${\bf x_e}$, to be suppressed at a later stage), involving (as
parameters) the nuclear coordinate set $\{Xr\}$. For any given value of $\{Xr\}$,
the $\zeta_k$ form a complete set, known as the adiabatic set or Born-Oppenheimer
electronic states, labeled by $k=1..\infty$. Let us consider only the finite set with
$k=1,…,n$ (where $n$ is in practice $2-4$). We shall call this the $P$-set and
designate its complement the $Q$ set. (This nomenclature follows previous
practice to decompose the full Hilbert space into two disjoint subspaces, $P$
and $Q$ \cite{HF}. Then the projection operator ${\bf I}$ over the full Hilbert space can be
written as the sum of two projection operators in the sub- spaces, namely
 ${\bf I}  = {\bf P} + {\bf Q}$ )
 The challenge has been to find a so-called diabatic set
\beq
\xi_m = \xi_m({\bf x_e},X_r)
\label{ksi}
\enq
related to the $P$ adiabatic-set alone, such that the nuclear derivative term in
the nuclear part of the Born-Oppenheimer equation corresponding to $\xi_m$
vanishes.
Baer \cite{MB1} showed that this can be achieved with an ADT matrix $A_{ij} (X_r)$ that
satisfies
\beq
A_{ij,r} + \tau^{(r)}_{ik} A_{kj}=0
\label{Aij}
\enq
for $i,j,k$ in $P$ and  all nuclear coordinate indexes $r$. (A tensor-algebra notation is
used here, but supplementing the double-index summation convention with
specification of the range of summands. The symbol $r$ after a comma
represents, as usual in tensor algebra, differentiation with respect to $X_r$).
Then
\beq
\xi_m = \zeta_k A_{km}
\label{xizeta}
\enq
($k,m$ in $P$) and the nuclear Schr\"{o}dinger equation contains no first-order
derivatives of the nuclear co-factors of $\xi_k$ . In \er{Aij} the following integrals over
the electronic coordinates appear:
\beq
\tau^{(r)}_{km} =<\zeta_k|\zeta_{m,r}>
\label{xizeta1}
\enq
\beq
\tau^{(r)}_{km} = -\tau^{(r)}_{mk}
\label{xizeta2}
\enq
The initial value conditions for $A_{km}$ for some chosen initial $\{X_r = X_{r0}\}$ are
\beq
A_{km}(\{X_{r0}\})=\delta_{km}
\label{Ainitial}
\enq                        
This is a convenient set of initial conditions, which can, however,
be generalized to cases when initially $A_{km}$ is non-zero for any $k$ and $m$ both inside
 $P$, or both inside $Q$ \cite {MB0}.
Reference \cite{MB1} also noted the following compatibility (curl) conditions as a
requirement for the solution of \er{Aij}:
\beq
\tau^{(r)}_{km,s} - \tau^{(s)}_{km,r} =
\tau^{(r)}_{kn}\tau^{(s)}_{nm} - \tau^{(s)}_{kn}\tau^{(r)}_{nm}
\label{tuacons}
\enq
($k,m,n$ in $P$; all unequal $r,s$. A shorthand way of writing this equation in terms
of many-dimensional vector matrices is $curl {\bf \tau} = - {\bf \tau x \tau}$.)
These relations arise
from differentiation of \er{Aij} with respect to $X_s$, a further differentiation with
respect to $X_r$ of another equation of the form \er{Aij} but involving $s$ instead of $r$,
and subtraction.
The actual requirement of compatibility is that
\beq
(\tau^{(r)}_{km,s} - \tau^{(s)}_{km,r}) A_{mh} =
(\tau^{(r)}_{kn}\tau^{(s)}_{nm} - \tau^{(s)}_{kn}\tau^{(r)}_{nm}) A_{mh}
\label{tuacons0}
\enq
($k,m,n,h$ in $P$). Satisfaction of the relations in \er{tuacons}
 is a sufficient condition for \er{tuacons0} to
hold for all and any $A_{mh}$ and for the existence of solutions in \er{Aij}.

 Having now formulated the general issue,we work out in the next section
 a case, involving Mathieu functions, in which \er {tuacons} is not satisfied.
These functions have already been used in the
context of the ADT matrix \cite {BYE}. We show here how the compatibility
condition can be better  and better satisfied by
successively enlarging in a systematic way the finite subset (the $P$
 subspace).
 Having  illustrated with the Mathieu functions the case of small
non-adiabatic coupling between $P$ and $Q$ subspaces, we then return
in section \ref {ADT} to the general theory and show what modifications are
needed in the ADT matrix, generally, when the non-adiabatic coupling between
$P$ and $Q$ are small, of order $ {\epsilon}$.
In section \ref {Analogy} we employ an analogy to show that, as in the case
of the ADT matrix, frequently a large Hilbert-space is needed {\it formally} , but in
practice a restricted space suffices.
\section {Mathieu Functions as Adiabatic States}
\label{MFAES}

    An electronic Schr\"{o}dinger equation for the angular electronic
coordinate $\theta$ and the polar nuclear coordinates $q$
and $\phi$ was written in Ref. \cite{BYE},
Eq. (1), as
\beq
[- \frac{1}{2} E_{el} \frac{\partial^2}{\partial \theta^2} -
G(q) cos(2 \theta -\phi ) - u(q, \phi)] \Psi (\theta |q,\phi) = 0
\label{Mathieweq}
\enq
where $E_{el}$ is a characteristic electronic quantity, $G(q)$
 is a nuclear-electronic
interaction coefficient, frequently assumed to be proportional to
 the nuclear
coordinate $q$ and $u(q, \phi)$ is the eigen-energy of the solution
 (being part of the
adiabatic potential for the nuclear motion).
  Equation (\ref{Mathieweq}) is recognized as the
Mathieu differential equation, which has enjoyed a wide literature and a
variety of notations \cite{WW}-\cite{MF}. We shall use
 the solutions given by these sources
(thus differing from the methodology in Ref. \cite{BYE},
in which the solutions were
derived), but shall retain the symbols for the parameters
 introduced in Ref. \cite{BYE}.
However, we shall make life easier for the reader by giving
 the relations
between the parameters. In the relations, we indicate the
 literature sources by
adding as subscripts the initials of the authors surnames.
In the spirit of Born-Oppenheimer approximation the small parameter in
the theory is
\beq
x \equiv \frac{G(q)}{E_{el}} =\frac{kq}{E_{el}} = 8 q_{_{WW}} = -q_{_{ML}} = -\frac{h^2_{_{MF}}}{4}
\label{xdef}
\enq
and the adiabatic angular coordinate for the electronic motion relative to the
nuclear one is
\beq
z \equiv \theta - \frac{\phi}{2}
\label{zdef}
\enq

The last relation implies that all derivatives with respect to $\phi$ can be replaced
by -$\frac{1}{2}$ times the derivative with respect to $z$.
In the notation of Ref. \cite{MAC}, the two families of solutions of \er{Mathieweq} that are
of interest to us have the form:
\beq
ce_{2n+1}(z,-x)= \sum_{r=0}^{\infty} A^{2n+1}_{2r+1} (-x) cos (2r+1) z  \qquad   (n=0,1,..)
\label{cedef}
\enq
\beq
se_{2n+1}(z,-x)= \sum_{r=0}^{\infty} B^{2n+1}_{2r+1} (-x) sin (2r+1) z  \qquad   (n=0,1,..)
\label{sedef}
\enq
The functions are conventionally normalized to $\pi$, so that the squares of the
coefficients sum up to $1$. Thus, when we use \eqs (\ref{cedef}
 - \ref{sedef}) for the wave-functions $\Psi$ of \er{Mathieweq} (that are normalized
in the square to unity) each function has to be
divided by $\sqrt{\pi}$. When we number the adiabatic electronic wave functions
according to the order of their energy surfaces  (as is commonly done
in molecular physics) by ($m'$), we have the correspondences (for $x>0$):
\ber
&&\Psi^{(1')} [\equiv \Psi^{(1)} (\theta |q,\phi)] = \frac{ce_{1}(z,-x)}{\sqrt{\pi}},
\Psi^{(2')} = \frac{se_{1}(z,-x)}{\sqrt{\pi}},
\nonumber \\
&&\Psi^{(3')}=\frac{ce_{3}(z,-x)}{\sqrt{\pi}},
\Psi^{(4')}=\frac{se_{3}(z,-x)}{\sqrt{\pi}},
\Psi^{(5')}= \frac{ce_{5}(z,-x)}{\sqrt{\pi}}, ...
\label{psi'}
\enr
 Solutions of \er{Mathieweq} of the form $ce_{2n}$, $se_{2n}$ have vanishing matrix elements with
those in \er{psi'} and can be disregarded. For small $|x|$, the following are the
leading terms in the expansions of the Mathieu functions for $n=0$ \cite{MAC} :
\beq
ce_{1}(z,-x)= (1-\frac{x^2}{128}) cos(z) + \frac{x}{8}(1- \frac{x}{8})cos(3z) +\frac{x^2}{192} cos(5z) +...
\label{ce1}
\enq
\beq
se_{1}(z,-x)= (1-\frac{x^2}{128}) sin(z) + \frac{x}{8}(1+ \frac{x}{8}) sin(3z) + \frac{x^2}{192} sin(5z) + ...
\label{se1}
\enq
Still for small $|x|$, the following are the leading terms in $A^{2n+1}_{2r+1} (-x)$ and
$B^{2n+1}_{2r+1} (-x)$ \cite{WW,MAC,DAV}:
for $r \ge n$:
\beq
\frac{(2n+1)!}{[(r-n)!(r+n+1)!]} (\frac{x}{4})^{r-n} +O(x^{r-n})
\label{rbign}
\enq
for $r<n$
\beq
\frac{(n-r)!}{[(n-r)! (2n)!]} (-\frac{x}{4})^{n-r} +O(x^{n-r})
\label{rsmn}
\enq
Now the eigenvalues $u$ of \er{Mathieweq} are of the
 order of $\frac{(2n+1)^2 E_{el}}{2}$
\cite{BYE,MAC,MF} (this can be seen from the leading
 terms of \er{cedef} and
\er{sedef} in which \er{rbign} and \er{rsmn} were
taken into consideration),
therefore the $n=0$ pair is well separated energetically from
the $n>0$  states that
lie higher. Adopting the nomenclature of section \ref{BaI}, we
shall call the two states
in \er{cedef} and \er{sedef} with $n=0$ the $P$ subset.
For $x=0$, $P$ is doubly degenerate; for
small $|x|$ it is nearly degenerate. We look at the nonadiabatic
 coupling matrices
$\tau$ within this subset. Using the definitions shown in
 \er{xizeta1} for $r=q$
and $\phi$ (see also equation (15) in Ref. \cite{BYE})
and the expansion in \er{ce1} and \er{se1} we obtain:
\beq
\tau^{(q)}_{1'2'} = 0
\label{tauq12}
\enq
and
\beq
\tau^{(\phi)}_{1'2'} = -\frac{(1+\frac{x^2}{32})}{2q}
\label{tauphi12}
\enq
agreeing with \eq (16) in Ref.\cite{BYE}. Turning now to our \er{tuacons},
 we write the
curl-term on the left hand side in the present curvilinear, plane-polar
coordinate system as:
\beq
\frac{1}{q} [\frac{\partial(q \tau^{(\phi)})}{\partial q}
- \frac{\partial(\tau^{(q)})}{\partial \phi}] = -\frac{x^2}{32 q^2}
= -\frac{k^2}{32 E_{el}^2}
\label{tuacons12}
\enq
where we have used \er{xdef} and \eqs ({\ref{tauphi12},\ref{tauq12})
in the last result.

However, evaluation of the right hand side of \er{tuacons}, ${\bf-\tau \times \tau}$,
 inside the $P$-set ($1',2'$)
shows that this is zero. Clearly, the missing part comes from the $P$-$Q$ inter-set
nonadiabatic coupling terms $\tau^{(r)}_{J \alpha}$.
When we evaluate ${\bf-\tau \times \tau}$ to the lowest order
in $x$, we find that this exactly matches the curl-value and that the missing
value come from the $n=1$ functions $ce_3$ and $se_3$. This means that enlarging the
$P$ set from $n=0$ to $n=0,1$ ensures the compatibility conditions for the solution of
the ADT matrix, correct to the order of $E_{el}^{-2}$.
To satisfy the curl-conditions to
higher powers in $E_{el}^{-1}$ (or $x$)
 will require bringing in higher $n$ values. It will be
presently shown that each higher power of ${x^2}$ requires one further $n$.

Table \ref{ABSX} shows the leading values for small $|x|$, as obtained from
 \er{rbign} and \er{rsmn}.

\begin{table}

\begin{tabular}{|c|c|c|c|c|} \hline\hline
$r/n$  & $0$ & $1$  & $2$ & $3$ \\ \hline\hline
$0$  & $1$ & $\frac{x}{8}$ & $O(x^2)$ & $O(x^3)$ \\ \hline
$1$  & $-\frac{x}{8}$ & $1$ & $\frac{x}{16}$ & $O(x^2)$ \\ \hline
 \end{tabular}
\caption{Some leading terms in the coefficients
$A$ and $B$ for small $|x|$}
\label{ABSX}
\end{table}

We now turn to the nonadiabatic couplings $\tau^{(r)}(J,\alpha)=
-\tau^{(r)}(\alpha,J)$, where $J$ is one
of the $n=0$ and $\alpha$ is one of $n>0$. Since we calculate these correct only to $x$, we
find from table \ref{ABSX},
upon recalling the power series expansions \er{ce1} and \er{se1}
for $ce_1$ and $se_1$ and the orthogonality of the trigonometric functions, that only
$n=1$ contributes nonvanishingly.
 We now draw up the list of the nonadiabatic
couplings labeled, express them in terms of quantum mechanical bra-kets of
the derivatives and give their values. We give those values of $r$ in the
expansions \er{cedef} and \er{sedef}, which contribute.
\ber
\tau^{(q)}_{1'3'} = & <ce_1|\frac{\partial}{\partial q} |ce_3>
\approx -\frac{x}{8q}  &(r=0)
\nonumber \\
\tau^{(\phi)}_{1'4'} = &-\frac{1}{2} <ce_1|\frac{1}{q}
\frac{\partial}{\partial z} |se_3>
\approx -\frac{x}{8q}  &(r=0,1)
\nonumber \\
\tau^{(q)}_{2'4'} = &<se_1|\frac{\partial}{\partial q}| se_3>
\approx -\frac{x}{8q}  &(r=1)
\nonumber \\
\tau^{(\phi)}_{2'3'} = & -\frac{1}{2} < ce_{2n+1}|\frac{1}{q}
\frac{\partial}{\partial z}|se_1>
\approx \frac{x}{8q}  &(r=0,1)
\label{hightau}
\enr
From these we calculate the vectorial cross-product in the ground
 [$P=(1',2')$]
doublet
\beq
({\bf \tau \times \tau})_{1'2'}=
\sum_{\alpha \in Q} [\tau^{(q)} (ce_1,\alpha) \tau^{(\phi)} (\alpha,se_1)
-\tau^{(\phi)} (ce_1,\alpha) \tau^{(q)} (\alpha,se_1)]
\label{tauctau}
\enq
Collecting all matrix elements, after noting the antisymmetric
 character of the
coupling matrices and the negative sign before second term
in \er{tauctau}, one finds
finally
\beq
({\bf \tau \times \tau})_{1'2'} = \frac{x^2}{32 q^2}
\label{tauctau2}
\enq
This agrees with the negative of the curl, as required. The contributing
intermediate states are seen to be the lowest energy set
 from $Q$ ($n=1$), which
is energetically immediately above the $P$ ($n=0$) set.
 This balances the $x^2$ term
in the curl, \er{tuacons12}. More generally, as one goes in $ce_1$, $se_1$
 (or in $\Phi^{(1')}$,$\Phi^{(2')}$) to
higher approximations in $x$, $curl \tau_{1'2'}$ will contain
 higher powers of $x^2$.
(Recall that $x \propto \frac{1}{E_{el}}$
is the small parameter in the Born-Oppenheimer approximation.)
We wish to show now that each consecutive term of order $x^{2s}$ in the
expansion of $curl \tau_{1'2'}$ will be balanced by a further set
 of the higher energy
wave functions, precisely up to $ce_{2s+1}$, $se_{2s+1}$
 and none higher. (Since the
method of computing the matrices is laborious,
 we do not calculate the higher
order terms in the $curl$ or in the vector
 product $({\bf \tau \times \tau})_{1'2'}$,
 only show from where the latter arise.)
The proof relies on the substitution of the dominant term,
 given in \er{rbign}
and \er{rsmn}, into the expansions (\ref{cedef})-(\ref{sedef})
 for the Mathieu functions.
We thus have approximately, for small $|x|$,
\beq
ce_{2n+1}(z,-x)\approx \sum_{r=0}^{\infty} a^{2n+1}_{2r+1} x^{|n-r|}
cos (2r+1) z  \qquad   (n=0,1,..)
\label{cedefap}
\enq
\beq
se_{2n+1}(z,-x) \approx \sum_{r=0}^{\infty} b^{2n+1}_{2r+1} x^{|n-r|}
 sin (2r+1) z  \qquad   (n=0,1,..)
\label{sedefap}
\enq
where $a$ and $b$ are (in the leading order) numerical. Next we form the
nonadiabatic coupling vector-matrix ${\bf \tau}$,
 whose elements are proportional to:
\beq
<ce_1|\frac{\partial}{\partial q}| ce_{2n+1}>
\approx \frac{1}{q} \sum_{r=0}^{\infty} a^{1}_{2r+1}
 a^{2n+1}_{2r+1} |n-r| x^{|n-r|+r}
\label{taumat1}
\enq
\beq
<ce_1|\frac{1}{q}\frac{\partial}{\partial z}| se_{2n+1}>
\approx \frac{1}{q} \sum_{r=0}^{\infty} a^{1}_{2r+1}
 b^{2n+1}_{2r+1} (2r + 1) x^{|n-r|+r}
\label{taumat2}
\enq
and two further matrix elements of similar form for coupling with $se_1$.
 To compensate the term of order $x^{2s}$ in $curl \tau_{1'2'}$ by a
similar term in $({\bf \tau \times \tau})_{1'2'}$, we
require that
\beq
|n-r|+r=s
\label{nrs}
\enq
Two cases have to be considered: $n \ge r$ and $n<r$.
In the former, \er{nrs} gives $n=s$.
In the latter case, one gets from \er{nrs} after transposing
\beq
2r=n+s
\label{nrs2}
\enq
If $n$ were greater than $s$, then
\beq
2r=n+s < 2n
\label{nrs3}
\enq
contradicting the assumption that $n<r$. Thus, in all cases, only Mathieu
functions of up to order $n=s$ need to be added to the intermediate set to
compensate the $x^{2s}$ term in the $curl$;
 higher order functions start with higher
powers of $x$. This leads to a hierarchical extension of the $P$ set
 to ensure the
compatibility condition to successively better accuracy in the
 ground, ($1',2'$) set.

\section{The Projected ADT Matrix}
\label {ADT}
We return now to the general case embodied in \er{tuacons} and \er {tuacons0}.

It has been established (Ref. \cite{MB1}, Appendix 1) that if in \er{tuacons} the
summation over $n$ is extended to $P+Q$, i.e. the intermediate states run over
the complete set, then \er{tuacons} is satisfied for $\tau^{(r)}_{km}$
 defined by (\ref{xizeta1})-(\ref{xizeta2}).

We shall now consider cases in which all $\tau^{(r)}_{km}$ are small,
(say) of order $\epsilon$,
for either $k$ or $m$ being in the set $Q$ complementary to $P$, though they can be large
for both $k$ and $m$ being in $P$ or in $Q$. Physical conditions for the existence of
 such situations have been noted before as either large energy gaps between
$P$ and $Q$ (\cite{PCK}, section III.A, end), or a strong vibronic coupling
localized in the neighborhood of some reference configuration  (\cite{PCK}, section III.B),
with which the $Q$-set has small overlap. In section 4 we consider yet another
situation, where the smallness of  $\epsilon$ arises from the small ratio
 between electronic and nuclear masses.

 Our aim is to show that \er{Aij}and \er{xizeta} continue
to be approximately valid, even though \er{tuacons} is not formally satisfied. Our
demonstration should disarm objections to some practical uses of the ADT
matrix. It also extends a previous proof \cite{MB2}, that the diabatic set exists when the
above considered $\tau^{(r)}_{km}$ are exactly zero, to those more frequently
encountered situations, where the non adiabatic couplings are non-zero, but
small.

Notation: Henceforth we shall use Latin subscripts (e.g.,$j, k, m, ..$) for the total
Hilbert space, capital  Latin subscripts (e.g., $J, K, M,..$ ) for the $P$ subset, and
Greek subscripts (e.g.,$\alpha, \beta,..$) for the $Q$ subset.
In this notation $\tau^{(r)}_{J \alpha}$ are all of
order $ \epsilon $ and small.
We now write out equations (\ref{Aij})
with initial conditions (\ref{Ainitial}) for the full
Hilbert space. Solutions of these exist, since the curl-condition in (\ref{tuacons}) is
satisfied for the full Hilbert-space. Explicitly, (\ref{Aij}) reads as follows:
\beq    
A_{JK,r} +\tau^{(r)}_{JM} A_{MK} + \tau^{(r)}_{J \alpha} A_{\alpha K} = 0
\label{AP}
\enq
for derivatives fully within $P$, and
\beq
A_{\alpha K,r} +\tau^{(r)}_{\alpha M} A_{MK}  + \tau^{(r)}_{\alpha \beta} A_{\beta K} = 0
\label{APQ}
\enq
for the inter - $PQ$ matrix element derivatives, and a further equation (which
will not be of interest to us) for the intra-$Q$ derivative matrix elements. Noting
that at $\{ X_r \} = \{ X_{r0} \}$ the initial conditions (\ref{Ainitial}) hold,
we assume now that the matrix
elements at a general point  $\{ X_r \}$ close to it can be expanded in powers of
$\{ X_r - X_{r0} \}$ as a Taylor series. The condition for this assumption is that
the matrix elements $\tau^{(r)}_{km}$ have no singularities in the neighborhood
considered \cite{MB0}. The expansion takes the following form:
\ber
&&A_{jk}(\{ X_r \})=
\sum_{N=0}^{\infty} \frac{1}{N!}[\sum_{r_{1} =1}^{M}  ...\sum_{r_{N}=1}^{M}
(X_{r_1} - X_{{r_1}0})...(X_{r_N} - X_{{r_N}0})
\nonumber \\
&& A_{jk,r_{1}...r_{N}}(\{ X_{r0} \})]
\label{Ataylor}
\enr
 in which the coefficients contain the derivatives $A_{jk,r_{1}...r_{N}}(\{ X_{r0} \})$
to the $N$'th order
($r_{1}...r_{N}$ contains $N$ similar or dissimilar symbols referring to the nuclear
coordinates, and $M$ is the number of nuclear coordinates).
We shall find that all inter-$PQ$ $A_{\alpha J}$ are of order $\epsilon$, but  the
corrections to the intra-$P$ $A_{JK}$ are merely of order $ \epsilon^2 $.

To start, we prove this for the first
derivatives $A_{JK,r} (\{ X_{r0} \})$ and \\
$A_{\alpha K,r} (\{ X_{r0} \})$ respectively.
These can be immediately evaluated from \er{AP}
 and \er{APQ}, with the following results:
 In \er{AP}, at $\{X= Xr0\}$ the third term is zero by the initial conditions (\ref{Ainitial})
$A_{km}(\{X_{t0}\})=\delta_{km}$, and
this gives the second term only, which has the same result as ignoring the $Q$ set, that is:
\ber    
&& A_{JK,r} (\{ X_{t0} \}) + \tau^{(r)}_{JM} (\{ X_{t0} \}) A_{MK} (\{ X_{t0} \}) =
\\ \nonumber
&&A_{JK,r} (\{ X_{t0} \}) + \tau^{(r)}_{JK} (\{ X_{t0} \}) = 0.
\label{APinitial}
\enr
In \er{APQ} the third term is similarly zero by the initial conditions, thus:
\ber
&& A_{\alpha K,r} (\{ X_{t0} \}) +\tau^{(r)}_{\alpha M} (\{ X_{t0} \}) A_{MK} (\{ X_{t0} \}) =
\\ \nonumber
&& A_{\alpha K,r} (\{ X_{t0} \}) +\tau^{(r)}_{\alpha K} (\{ X_{t0} \}) = 0
\label{APQinitial}
\enr
we conclude that $A_{\alpha K,r} (\{ X_{t0} \}) = -\tau^{(r)}_{\alpha K} (\{ X_{t0} \})$
is of the same order of $ \epsilon $, due to $\tau^{(r)}_{\alpha K} (\{ X_{t0} \})$ being of this order.

In the appendix we establish by mathematical induction that the corrections to all
the inter-$PQ$ set derivatives $A_{\alpha K,r_{1}...r_{N}}(\{ X_{t0} \})$
 are of order $\epsilon$, and that the
corrections to the intra-P derivatives $A_{JK,r_{1}...r_{N}}(\{ X_{t0} \})$
 are of order $\epsilon^2$. Subsequent derivatives are obtained recursively,
rather than through integration, so that the validity of the procedure depends on
the convergence of the Taylor series (which is assumed to hold close to the
initial point).

Summarizing, subject to the assumption of small (though not necessarily
zero) inter-$PQ$ non-adiabatic coupling and analyticity of the matrix elements,
the original procedure of  Ref. \cite{MB1} restricted to a finite dimensional set ($P$) is
approximately valid, and correct to order $\epsilon^2$, no matter that the compatibility
conditions are not exactly satisfied. Earlier, it was shown, for the model
discussed here, that the discrepancy in the compatibility conditions
 (\ref{tuacons}) is $\epsilon^2$
(e.g., Ref. \cite{PCK,TK}); our result establishes the magnitude of error in the solution.

\section {An Inductive Proof}

We now establish by mathematical induction that the corrections to all
the inter-$PQ$ set derivatives $A_{\alpha K,r_{1}...r_{N}}(\{ X_{t0} \})$
are of order $\epsilon$, and that the
corrections to the intra-$P$ derivatives
$A_{JK,r_{1}...r_{N}}(\{ X_{t0} \})$
are of order $\epsilon^2$.

This result was already proven for $N=1$. We assume that the result is correct for
$N-1$ and show that the result is also correct for $N$. We can write:
\beq
A_{\alpha K,r_{1}...r_{N}}(\{ X_{t0} \})=
\partial_{r_{1}...r_{m-1} r_{m+1}...r_{N}} A_{\alpha K,r_{m}}(\{ X_{t0} \})
\label{induc1}
\enq
Using \er{APQ} this can be written also as:
\beq
A_{\alpha K,r_{1}...r_{N}}(\{ X_{t0} \})=
-\partial_{r_{1}...r_{m-1} r_{m+1}...r_{N}}
(\tau^{(r_{m})}_{\alpha M} A_{MK}  + \tau^{(r_{m})}_{\alpha \beta} A_{\beta K})
(\{ X_{t0} \})
\label{induc2}
\enq
Or also as:
\ber
A_{\alpha K,r_{1}...r_{N}}(\{ X_{t0} \}) &=&
-\partial_{r_{1}...r_{m-1} r_{m+1}...r_{N}}
(\tau^{(r_{m})}_{\alpha M} A_{MK}) (\{ X_{t0} \})
\nonumber \\
&-&\partial_{r_{1}...r_{m-1} r_{m+1}...r_{N}}
(\tau^{(r_{m})}_{\alpha \beta} A_{\beta K})
(\{ X_{t0} \})
\label{induc3}
\enr
The first term in the right hand side
of the above equation is proportional to $\tau^{(r_{m})}_{\alpha M}$
and its derivatives, and thus by assumption is of order $\epsilon$.
The second term in the right hand side
of the above equation contains derivatives up to order $N-1$ of $A_{\beta K}$
and thus by the assumption of the induction is also of order $\epsilon$.
Thus we established that the right hand side of \er{induc3} is of order $\epsilon$
and  also $A_{\alpha K,r_{1}...r_{N}}(\{ X_{t0} \})$ is of order $\epsilon$.
We conclude that $A_{\alpha K,r_{1}...r_{N}}(\{ X_{t0} \})$ is of order $\epsilon$
for all $N$ by induction.

Next we want to establish that correction to $A_{JK,r_{1}...r_{N}}(\{ X_{t0} \})$
are of order $\epsilon^2$. We established this result for $N=1$ inside the text.
We proceed by assuming that this result is true for $N-1$ and will prove its
correctness  for $N$, again we write:
\beq
A_{JK,r_{1}...r_{N}}(\{ X_{t0} \})=
\partial_{r_{1}...r_{m-1} r_{m+1}...r_{N}} A_{JK,r_{m}}(\{ X_{t0} \})
\label{induc4}
\enq
Using \er{AP} we obtain:
\beq
A_{JK,r_{1}...r_{N}}(\{ X_{t0} \})=
-\partial_{r_{1}...r_{m-1} r_{m+1}...r_{N}}
(\tau^{(r_{m})}_{J M} A_{MK}  + \tau^{(r_{m})}_{J \beta} A_{\beta K})
(\{ X_{t0} \})
\label{induc5}
\enq
Or also as:
\ber
A_{JK,r_{1}...r_{N}}(\{ X_{t0} \})&=&
-\partial_{r_{1}...r_{m-1} r_{m+1}...r_{N}}
(\tau^{(r_{m})}_{JM} A_{MK}) (\{ X_{t0} \})
\nonumber \\
&-&\partial_{r_{1}...r_{m-1} r_{m+1}...r_{N}}
(\tau^{(r_{m})}_{J\beta} A_{\beta K})
(\{ X_{t0} \})
\label{induc6}
\enr
The first term in the right hand side
of the above equation contains intra-P terms which are not considered corrections.
The second term in the right hand side
of the above equation,
which is considered a correction term,
contains derivatives up to order $N-1$ of $A_{\beta K}$,
which are of order $\epsilon$ according to the theorem established above.
When multiplied by derivatives of $\tau^{(r_{m})}_{\alpha M}$
which are of order $\epsilon$ by our assumptions,
the correction terms obtained is thus of order $\epsilon^2$ for N. Thus the
 result holds for any $N$, which is what we have set out to prove.

\section{An Analogy}
\label {Analogy}
Very frequently one has a complicated Hamiltonian $H$, which has non-
zero matrix elements $H_{nm}$ between the components (designated $n$ and $m$) in
an extended Hilbert space. For many practical purposes, e.g. for numerical
work, one uses a restricted, finite set (belonging to the $P$ subset of the Hilbert
space), such that $P$-$Q$ matrix elements of the Hamiltonian are small.  The
procedure of solving the Schr\"{o}dinger equation in this manner can be formally
written as:
\beq
H {\bf P}=E_{P} {\bf P}
\label{HP}
\enq
where, as before, ${\bf P}$ is a projection operator and $E_{P}$
 represents the set of
eigenvalues corresponding to the set of solutions thus obtained.
Pre-multiplication with ${\bf Q}$, gives
\beq
{\bf Q} H {\bf P}= E_{P} {\bf Q}  {\bf P} = 0
\label{QHP}
\enq
since $P$ and $Q$ are disjoint.
This implies that the procedure requires, for
consistency, that $ {\bf Q} H {\bf P}=0$, or that there be
 no matrix elements $H_{nm}$  of the
Hamiltonian connecting the $Q$ and $P$ sets.
 While this is, in general, a true
requirement for exact solutions, in practice, when one seeks meaningful
approximate results, one is content with the smallness of these matrix
elements. We have here a further example of a rigorous condition being
inconsistent with an approximate method.

\section{Conclusion}

    In section \ref{MFAES} we have seen that for adiabatic electronic
 wave functions in the form of Mathieu functions the compatibility (or curl) conditions are
not satisfied but, by successively
enlarging the finite set, the conditions can be satisfied to increasingly better
 accuracy. As previously shown in (e.g.) \cite {PCK} Eq. (2.54), a non-zero
 curl gives rise to a ''gauge field tensor, the non-Abelian analog of
 the electromagnetic field", also called the Yang-Mills field \cite{YM}.
  The use of
 Mathieu functions will thus enable a systematic study of the properties of
 this field in a molecular physical context \cite {MSW}.

We then discussed the question of the existence of a finite sized ADT
matrix $A$ (which is distinct from the question of its uniqueness or single
valuedness), being the solution of a finite number of partial differential
equations that fail to satisfy compatibility conditions. We have started with a
complete set, for which the existence of solution is assured. Under conditions
that the nonadiabatic coupling with the exterior of the set is small, we have
truncated the size of the set and removed part of the finite sized ADT matrix
$A$. The compatibility conditions are not met within the finite set, but the
remainder of $A$ satisfies the original differential equation approximately,
namely to an accuracy which goes as the square of the neglected
nonadiabatic coupling. We conclude that the failure of the compatibility
conditions yet permits the existence of an approximate ADT matrix (subject to
smallness of some nonadiabatic coupling terms).

\end{document}